\documentclass[preprint,12pt]{elsarticle}

\usepackage{amsmath}
\usepackage{amssymb} 
\usepackage{cases}

\newcommand{\x}[1]{{\bf x}_{#1}}

\journal{Optics Communications}

\begin{document}

\begin{frontmatter}



\title{Improvement of the image quality of random phase--free holography using an iterative method}


\author[a]{Tomoyoshi Shimobaba\corref{cor1}}
\cortext[cor1]{Tel: +81 43 290 3361; fax: +81 43 290 3361}
\ead{shimobaba@faculty.chiba-u.jp}
\author[a]{Takashi Kakue}
\author[a]{Yutaka Endo}
\author[a]{Ryuji Hirayama}
\author[a]{Daisuke Hiyama}
\author[a]{Satoki Hasegawa}
\author[a]{Yuki Nagahama}
\author[a]{Marie Sano}
\author[a]{Minoru Oikawa}
\author[a]{Takashige Sugie}
\author[a]{Tomoyoshi Ito}

\address[a]{Chiba University, Graduate School of Engineering, 1--33 Yayoi--cho, Inage--ku, Chiba, Japan, 263--8522}

\begin{abstract}
Our proposed method of random phase-free holography using virtual convergence light can obtain large reconstructed images exceeding the size of the hologram, without the assistance of random phase.
The reconstructed images have low-speckle noise in the amplitude and phase-only holograms (kinoforms); however, in low-resolution holograms, we obtain a degraded image quality compared to the original image.
We propose an iterative random phase-free method with virtual convergence light to address this problem.
\end{abstract}


\end{frontmatter}

\section{Introduction}
Digital holographic display is a promising technique,  because a wavefront of light scattered from an object can be appropriately reconstructed from the display; therefore,  this property will enable the realization of an ideal three-dimensional display and projector. 
Holographic projections \cite{hj1, hj2, hj3, hj4, hj5} have unique properties, including multi-projection \cite{multi} ( by which a multi-image is projected on multiple screens),   projection on screens of arbitrary surface, and lensless zoom-able holographic projection \cite{zoom1,zoom2,zoom3}.
The lensless zoom-able holographic projection will lead to the development of an ultra-small projector.
Reconstructed images exceeding the hologram size, in general, require the random phase; however, this causes considerable problems of speckle noise.

There are well-known methods for improving the random phase applied-holograms: the Gerchberg-Saxton (GS) algorithm \cite{ite1,ite2}, the multi-random phase method \cite{amako}, the one-step-phase-retrieval method (OSPR) \cite{ospr}, and pixel separation methods \cite{pix1,pix2,pix3}.
The multi-random phase and pixel separation methods require display devices with high-speed refresh rates.
Conversely, random phase-free methods have also been proposed, for example, the error diffusion \cite{ospr,ed1,ed2,ed3,ed4} and down-sampling methods \cite{down}.
These methods can reconstruct clear images; however, the size of the reconstructed image cannot exceed the size of the hologram, because light fron the object does not spread widely \cite{hj5}.
Therefore, these methods cannot be used for lensless zoom-able holographic projections.

Recently, new random phase-free methods using virtual special convergence light for amplitude and phase-only holograms (kinoforms) have been proposed \cite{rand_free1, rand_free2, rand_free3}. 
Without the assistance of the random phase, this method can reconstruct images that exceed the hologram's size with low-speckle noise; however, in low-resolution holograms, we  obtain a degraded image quality by ringing artifacts, as will be shown in the next section.

In this paper, we propose an iterative random phase-free method with virtual convergence light to address this problem.
Sections 2 and 3 describe the proposed method and the  results of simulation performed using it.
Section 4 concludes this work.

\section{Proposed method}

\begin{figure}[htb]
\centerline{
\includegraphics[width=10cm]{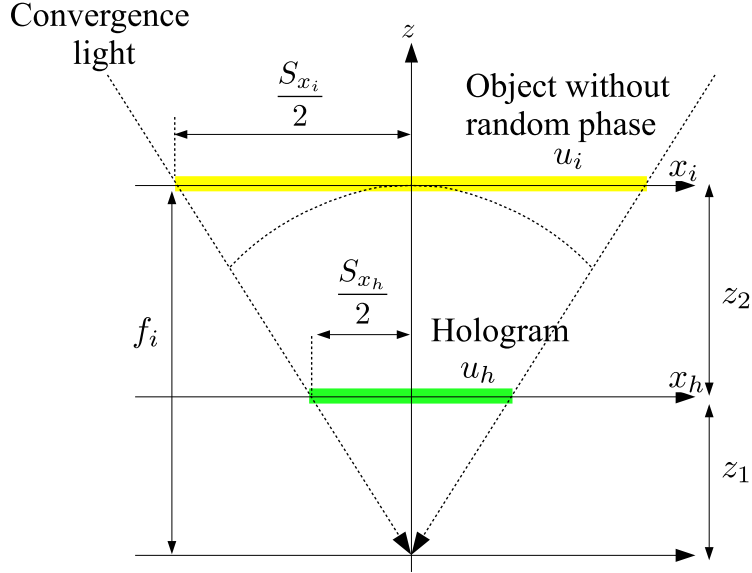}}
\caption{Calculation setup for the random phase-free method using virtual convergence light.}
\label{fig:setup}
\end{figure}

The random phase-free method using virtual convergence light is outlined in \cite{rand_free1, rand_free2, rand_free3}.
The calculation setup is shown in Fig. \ref{fig:setup}.
This method is applicable to amplitude computer-generated holograms (CGHs) and kinoforms.
Instead of using the random phase, the complex amplitude on the image plane $u_i(x_i,y_i)$ is multiplied using virtual convergence light given by 
\begin{equation}
w(x_i, y_i)=\exp(-i \pi(x_i^2+y_i^2)/\lambda f_i),
\end{equation}
where $f_i=z_1+z_2$ is the focal length,
$z_1$ is the distance between the focus point of the convergence light and the hologram, and is set to the distance at which the hologram just fits to the cone of the convergence light, and $z_2$ is the distance between the object and hologram.

Here we describe how to determine $f_i$.
Using a simple geometric relation, we can derive $S_{h}/2 : S_{i}/2 = z_1: f_i$ where the areas of the image and the amplitude CGH (or kinoform) are given by $S_{i} \times S_{i}$ and $S_{h} \times S_{h}$, respectively.
Therefore, we obtain:
\begin{equation}
f_i=z_2/(1-S_{h}/ S_{i}).
\end{equation}
To avoid overlap between the reconstructed image and the 0-th order light, the original object must be shifted from the optical axis by a distance of $o$.
Owing to the addition of this shift amount, the focal length of the convergence light is expressed as
\begin{equation}
f_i=z_2/(1-S_{h}/ (S_{i} + 2 o)).
\end{equation}

We calculate the complex amplitude in the hologram plane by
\begin{equation}
u_h(x_h, y_h) = {\rm Prop_{z_2}}\{u_i(x_i, y_i) w(x_i, y_i)\}, 
\end{equation}
where $ {\rm Prop_{z_2}\{\cdot \}}$ denotes the diffraction calculated at the propagation distance, $z_2$.
The following equation is used to calculate the amplitude CGH, $I(x_h,y_h)$ from $u_h(x_h, y_h)$:
\begin{equation}
I(x_h,y_h) = \Re{\{u_h(x_h, y_h)\}}, 
\end{equation}
where $\Re{\{\cdot\}}$ denotes the real part of $u_h(x_h, y_h)$.
In calculating the kinoform, $\theta(x_h,y_h)$, from $u_h(x_h, y_h)$, the following equation is used:
\begin{equation}
\theta(x_h, y_h) = {\rm arg} \{u_h(x_h, y_h) \},
\end{equation}
where ${\rm arg} \{\cdot \} $ indicate the use of only the argument of the complex amplitude. 
Unfortunately, in kinoforms, the reconstructed images using the random phase-free method are heavily degraded.
This is a common problem of kinoforms \cite{ed3,ed4,down}.
To overcome this problem, we apply the error diffusion method \cite{ed3,ed4} to the random phase-free kinoform as follows:
\begin{equation}
\theta(x_h, y_h) = {\rm ED} \{\theta(x_h, y_h) \},
\end{equation}
where $ {\rm ED} \{\cdot\}$ denotes the error diffusion operator.
The details of the error diffusion are described in Section \ref{sec:ite-kino}.
For the details of the random phase free-amplitude CGH and kinoform, refer Ref. \cite{rand_free1,rand_free2}.

\begin{figure}[htb]
\centerline{
\includegraphics[width=10cm]{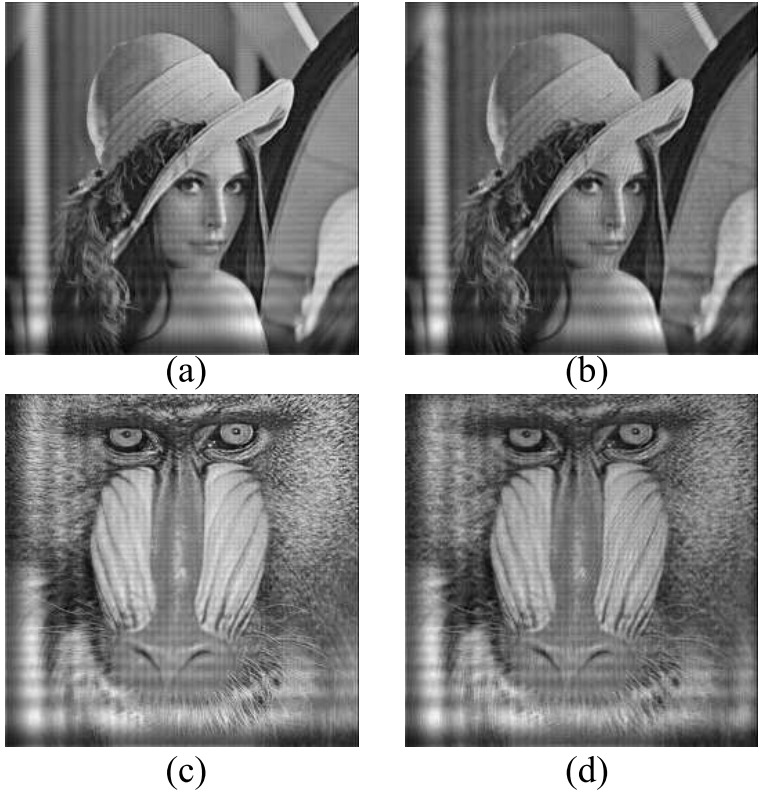}}
\caption{Reconstructed images from amplitude CGH and kinoform. (a) and (c) reconstructed images from amplitude CGHs: (b) and (d) reconstructed images from kinoform: These reconstructed images are contaminated by ringing artifacts.}
\label{fig:degrade}
\end{figure}

Figure \ref{fig:degrade} shows reconstructed images from the amplitude CGH and kinoform. 
Figures \ref{fig:degrade}(a) and \ref{fig:degrade}(c) are the reconstructions from the amplitude CGHs of the images ``Lena'' and ``Mandrill', respectively. 
Figures \ref{fig:degrade}(b) and \ref{fig:degrade}(d) are the reconstructions from the kinoforms of the images, respectively.
These reconstructed images do not have the speckle noise induced by the random phase; however, ringing artifacts overlap in these reconstructed images.

\subsection{Improved image quality using an iterative method}
\label{sec:ite-kino}

\begin{figure}[htb]
\centerline{
\includegraphics[width=12cm]{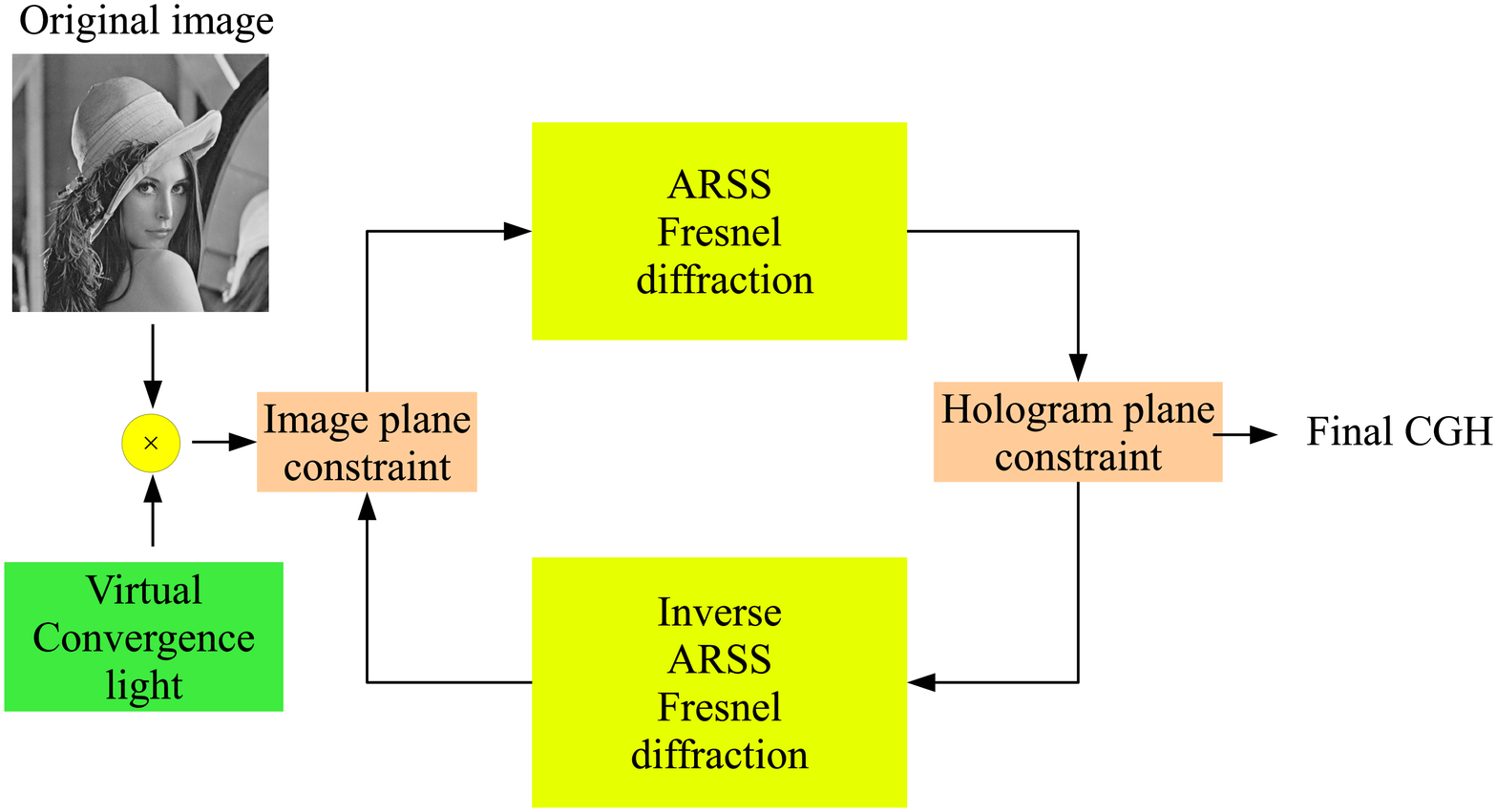}}
\caption{Iterative method for amplitude CGH.}
\label{fig:ite}
\end{figure}

\begin{figure}[htb]
\centerline{
\includegraphics[width=12cm]{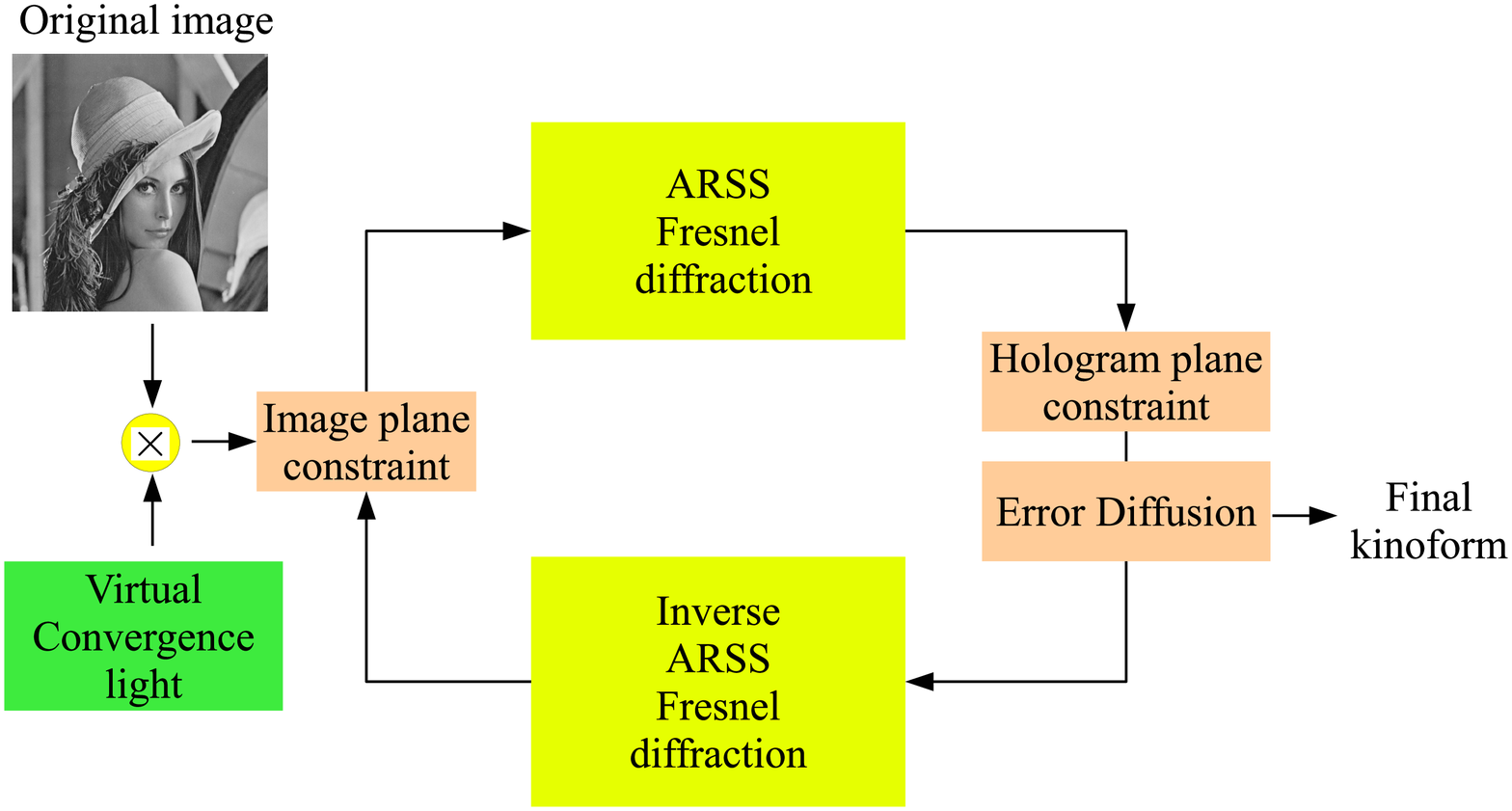}}
\caption{Iterative method for kinoform.}
\label{fig:ite-kino}
\end{figure}

To overcome the problem with ringing artifacts problem in the reconstructed images, we apply an iterative enhancement of the random phase-free method.
Figures \ref{fig:ite} and \ref{fig:ite-kino} show the improved method for amplitude CGHs and kinoforms using the iterative algorithm, respectively.
The kinoform yielded by the iteration method is almost the same as that in Fig. \ref{fig:ite}, except for the error diffusion.
The calculation steps are as follows:

\begin{enumerate}
\item We start the iteration by multiplying the virtual convergence light $w(x_i, y_i)$ by the original image $a_i(x_i, y_i)$.
The image plane is denoted by $u_i(x_i,y_i)= a_i(x_i, y_i) w(x_i, y_i)$. 

\item We calculate the scaled and shifted diffraction, which can be used to obtain light propagation at different sampling pitches on the image and hologram, as well as the off-axis propagation, by $u_h(x_h, y_h) = {\rm Prop_{z_2}}\{u_i(x_i, y_i) w(x_i, y_i)\}$.
The diffraction we used is ARSS Fresnel diffraction \cite{arss}, which is a scaled and shifted diffraction.
The sampling pitches on the image and hologram planes are set to $p_i$ and $p_h$, respectively.
The amount of shift away from the optical axis on the image plane is $(o_x,o_y)$.]

\item We extract only the real part of  $u_h(x_h, y_h)$ for the amplitude CGH or only the argument of $u_h(x_h, y_h)$ for the kinoform (which represents constraints in the hologram plane).
\begin{subnumcases}
{g(x_h,y_h)=}
\Re{\{u_h(x_h, y_h)\}} & (For amplitude CGH) \\
{\rm ED} \{ {\rm arg} \{u_h(x_h, y_h) \} \} & (For kinoform)
\end{subnumcases}

\item We back-propagate $u_h(x_h, y_h)$ to the image plane, $u_i'
(x_i, y_i) = {\rm Prop_{-z_2}}\{g(x_h, y_h)\}$ by ARSS Fresnel diffraction.
If we use the offset parameter, the shift amount away from the optical axis should be set to $(-o_x, -o_y)$.

\item We update only the amplitude of $u'_i(x_i, y_i)$ according to
\begin{equation}
u_i(x_i, y_i)=a_i(x_i, y_i) u'_i(x_i, y_i)/|u'_i(x_i, y_i)| .
\end{equation}
This is the constraint in the image plane.
Subsequently, we obtain the new image plane $u_i(x_i, y_i)$.
\end{enumerate}
We repeat the above iteration steps (2) to (5) until reaching the preset iteration number or a good image quality.

In the step 3, the Floyd-Steinberg error diffusion method we used was calculated using  
\begin{eqnarray}
\theta(x_h, y_h+1) &\leftarrow&  \theta(x_h, y_h+1) + w_1 e(x_h, y_h), \\
\theta(x_h+1, y_h-1) &\leftarrow&  \theta(x_h+1, y_h-1) + w_2 e(x_h, y_h), \\
\theta(x_h+1, y_h) &\leftarrow&  \theta(x_h+1, y_h)  + w_3 e(x_h, y_h), \\
\theta(x_h+1, y_h+1) &\leftarrow&  \theta(x_h+1, y_h+1) + w_4 e(x_h, y_h), 
\end{eqnarray}
where the coefficients are $w_1$=7/16, $w_2$=3/16, $w_3$=5/16, and $w_4$=1/16 \cite{ed3}: $\leftarrow$ indicates that the current value is overwritten, and $e(x_h, y_h)=u_h(x_h, y_h) -\theta(x_h, y_h)$.
Before applying the error diffusion, we may need to normalize the complex amplitude $u_h(x_h, y_h)$ by $u_h(x_h, y_h) \leftarrow u_h(x_h, y_h) / d$, where $d={\rm max}|u_h(x_h, y_h)|$ is the maximum absolute value in the kinoform. 

\section{Results}
We show the effectiveness of the proposed method through simulation.
The calculation is performed under conditions where the resolution of the images is $2,048 \times 2,048$ pixels, the wavelength is 532 nm, the pixel pitches on the image plane and the hologram plane are $p_i=8 \mu$m and $p_h=4 \mu$m, respectively, the distance between the image and hologram planes is 1 m, and the amount of shift away from the optical axis is $(o_x,o_y)=(20 {\rm mm}, 20 {\rm mm})$.

\begin{figure}[htb]
\centerline{
\includegraphics[width=14cm]{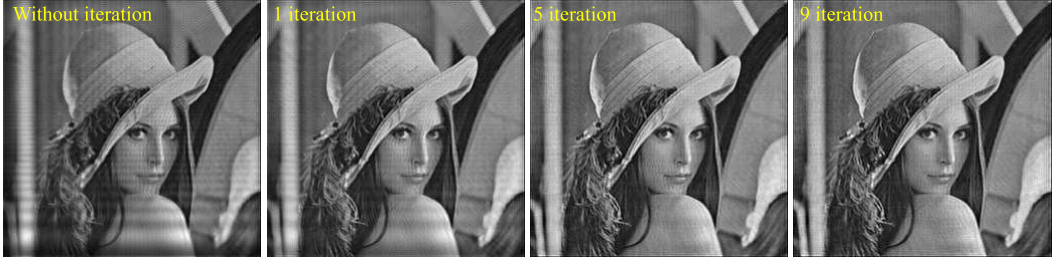}}
\caption{Reconstructions of the image ``Lena''. The upper and lower images are reconstructed from amplitude CGHs and kinoforms, respectively. They are generated by the proposed method.}
\label{fig:ite-result-lena}
\end{figure}

\begin{figure}[htb]
\centerline{
\includegraphics[width=12cm]{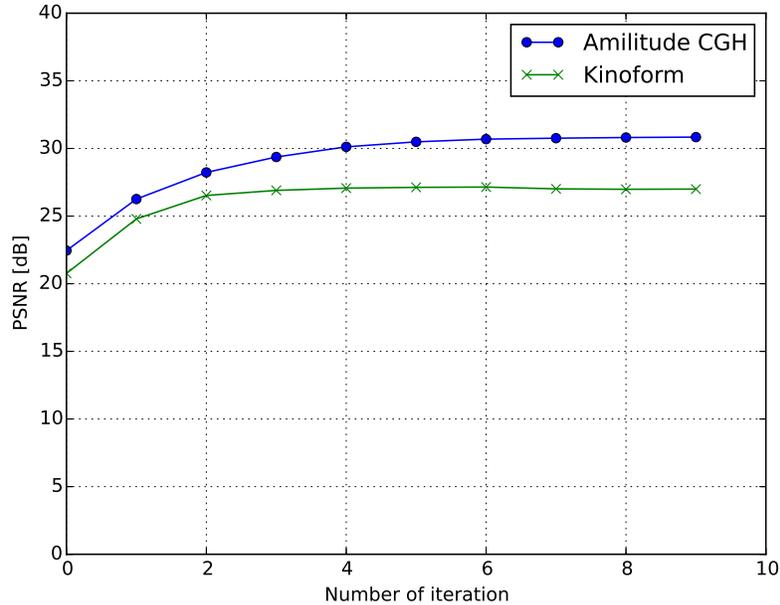}}
\caption{PSNRs for the reconstructed images of ``Lena'' from the amplitude CGHs and kinoforms.}
\label{fig:psnr-lena}
\end{figure}

The upper images in Fig. \ref{fig:ite-result-lena} show  reconstructions of the image ``Lena'' from amplitude CGHs, and the lower images show reconstructions kinoforms, which are generated by the proposed method.
Increasing the iteration number results in improvement of the image quality in both amplitude CGH and kinoform.
Figure \ref{fig:psnr-lena} shows the peak signal-to-noise ratio (PNSR) between the original image and the reconstructed image when increasing the number of iteration.
The blue and green lines indicate the PSNRs for the amplitude CGHs and kinoforms, respectively.
The improvement effect converges at five iterations, and the PSNRs of the amplitude CGH and kinoform are about 30 dB and 27 dB, respectively.

\begin{figure}[htb]
\centerline{
\includegraphics[width=14cm]{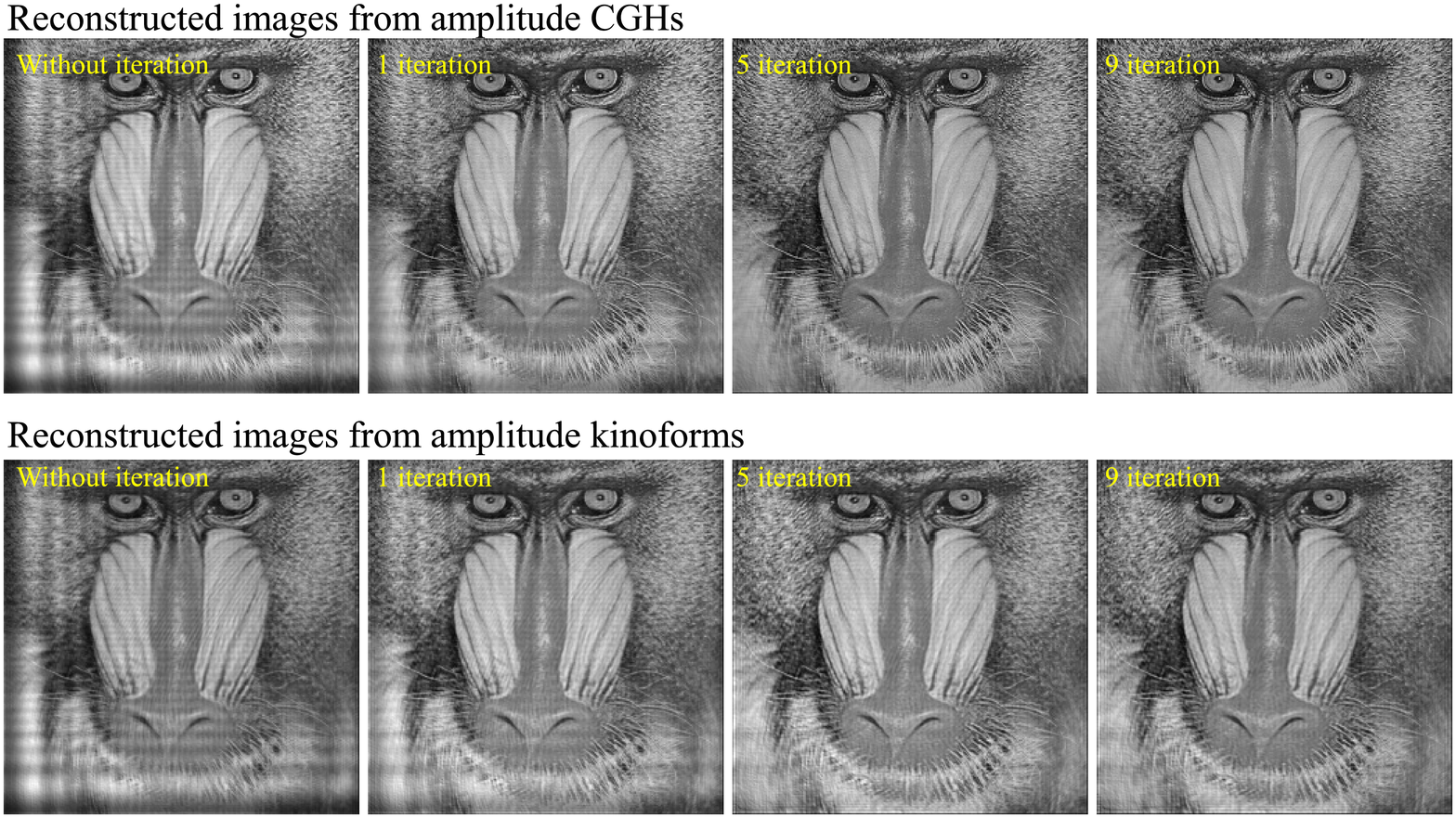}}
\caption{Reconstructions of the image ``Mandrill''. The upper and lower images are reconstructed from amplitude CGHs and kinoforms, respectively. They are generated by the proposed method.}
\label{fig:ite-kino-mandrill}
\end{figure}

\begin{figure}[htb]
\centerline{
\includegraphics[width=12cm]{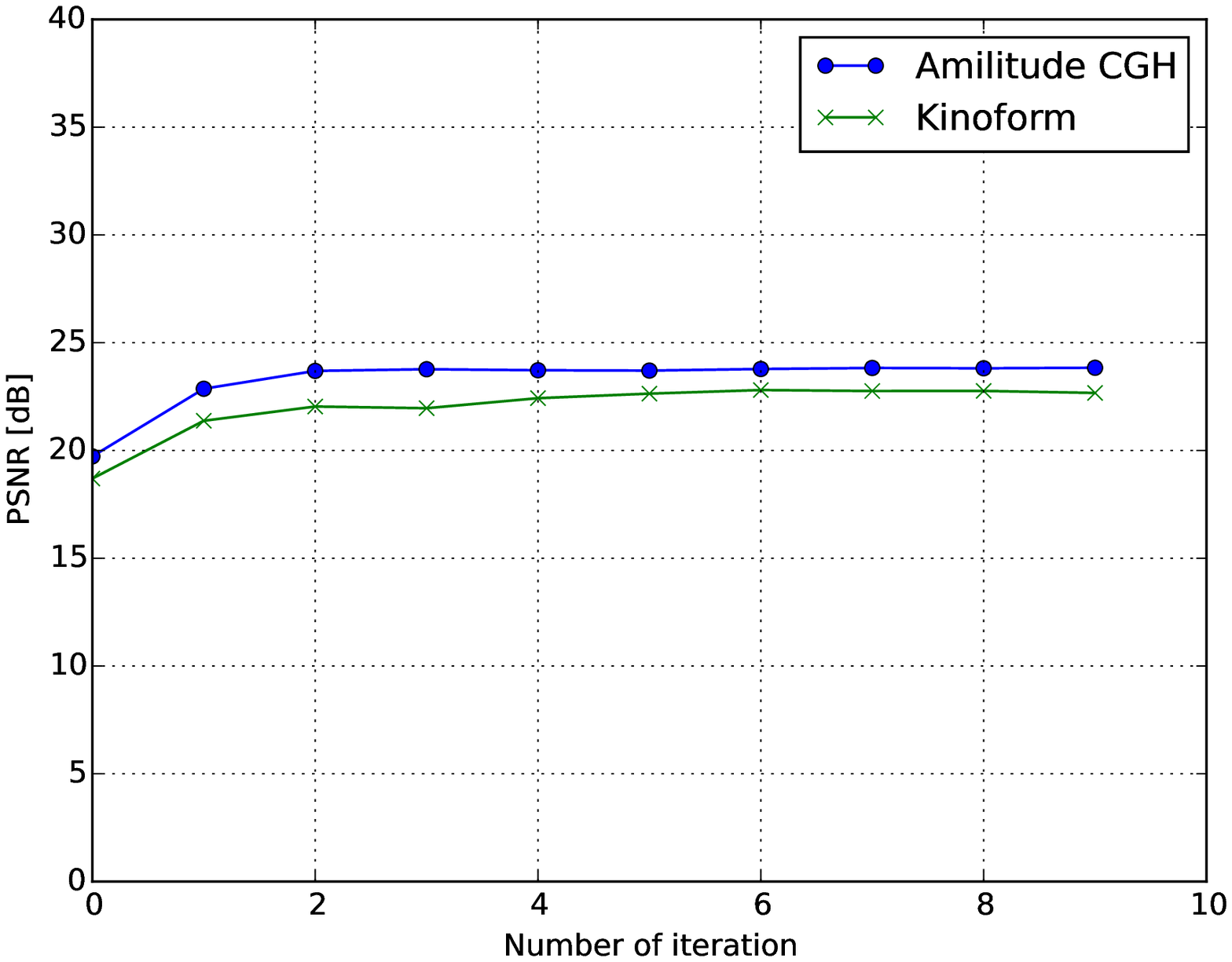}}
\caption{PSNRs for the reconstructed images of ``Mandrill'' from the amplitude CGHs and kinoforms.}
\label{fig:psnr-mandrill}
\end{figure}

Figure \ref{fig:ite-kino-mandrill} shows the reconstructions of the image ``Mandrill''.
The upper and lower images are obtained from amplitude CGHs and kinoforms generated by the proposed method, respectively.
Figure \ref{fig:psnr-mandrill} shows the PNSR between the original and the reconstructed images when increasing the number of iteration.
The improvement effect is the same as that in Fig.\ref{fig:psnr-lena}, and the PSNRs at five iterations of the amplitude CGH and kinoform are about 23 dB and 22 dB, respectively.
Overall, the proposed method greatly improves the image quality.

\begin{figure}[htb]
\centerline{
\includegraphics[width=14cm]{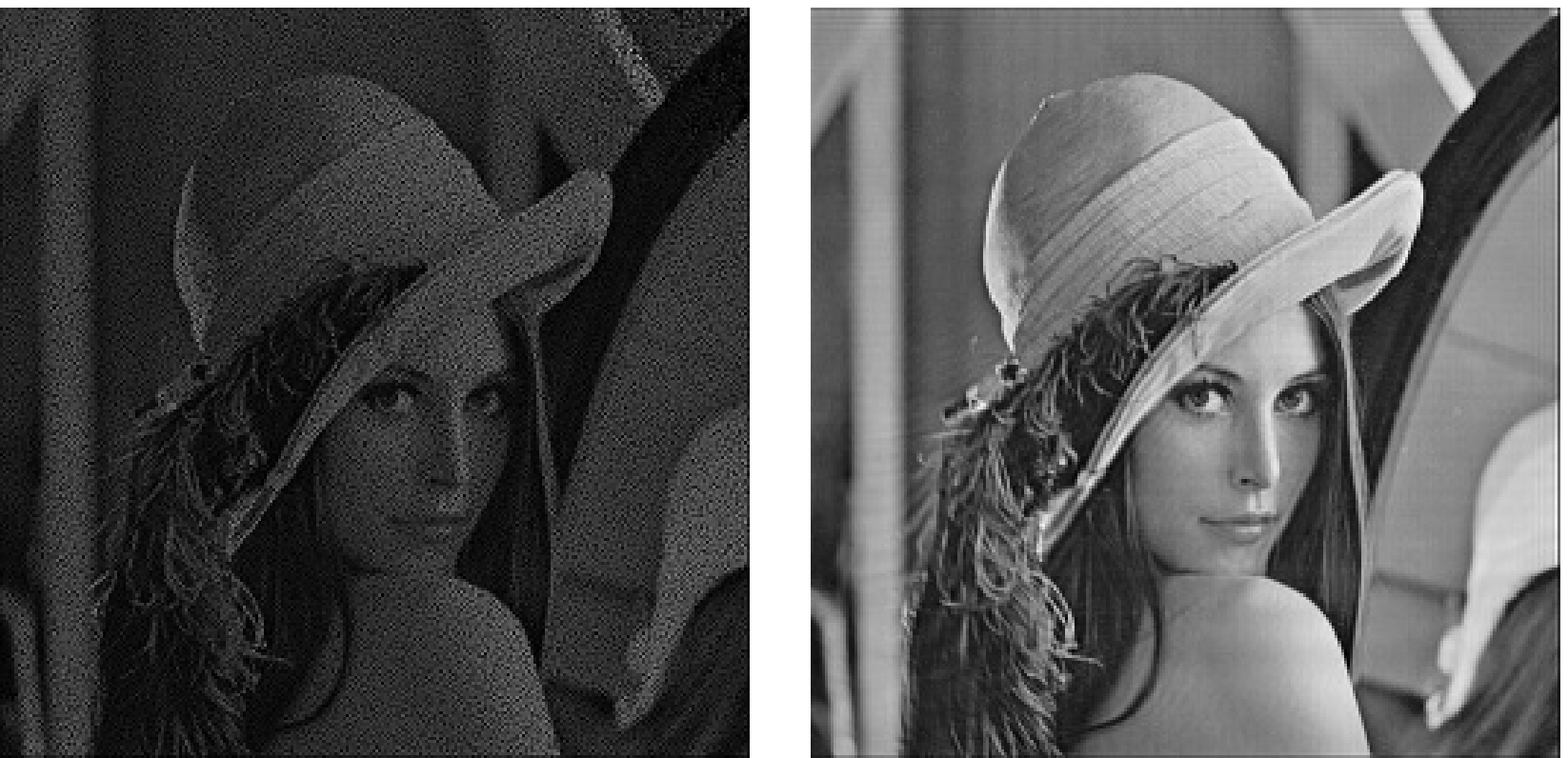}}
\caption{Reconstructed images at five iterations.  With random phase (left). With the proposed method (right).}
\label{fig:comp-amp}
\end{figure}

\begin{figure}[htb]
\centerline{
\includegraphics[width=14cm]{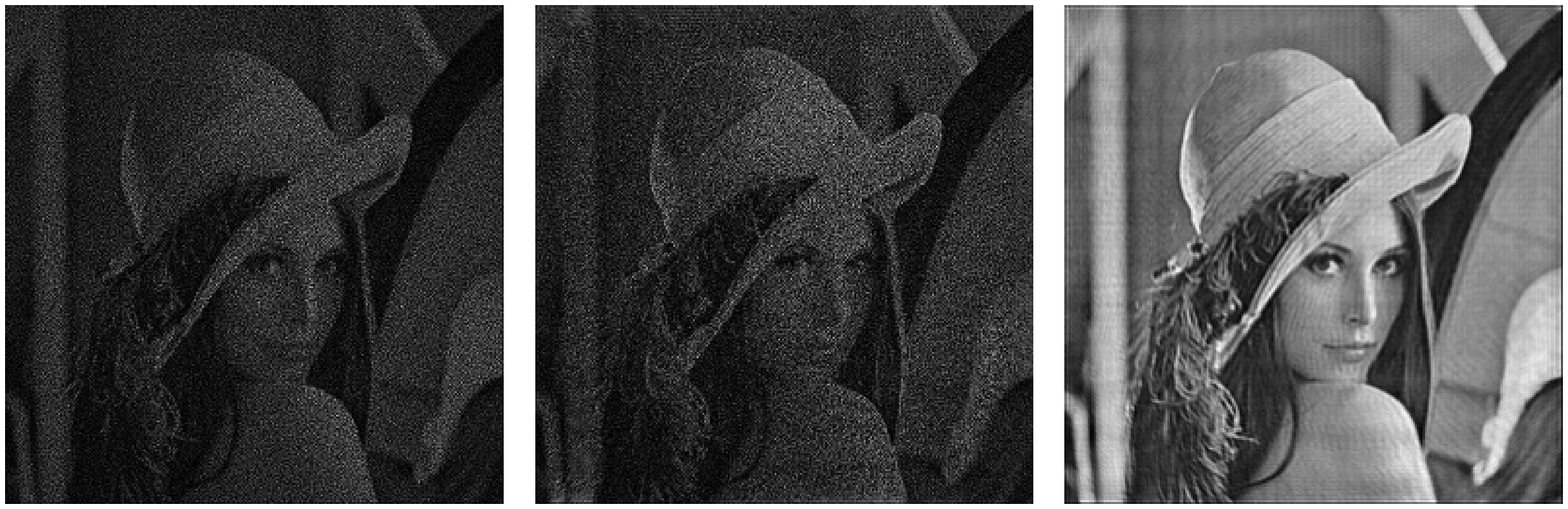}}
\caption{Reconstructed images at five iterations. With random phase (left). With the proposed method not applying the error diffusion (middle). With the proposed method using the error diffusion (right).}
\label{fig:comp-kino}
\end{figure}

Figure \ref{fig:comp-amp} shows the reconstructed images from amplitude CGHs with random phase (left) and with the proposed method (right) at five iterations.
The PSNRs for the left and right images are 9.8 dB and  30.5 dB, respectively.
Figure \ref{fig:comp-kino} shows the reconstructed images from kinoforms with random phase (left), with the proposed method not applying the error diffusion (middle) and with the proposed method using the error diffusion (right)  at five iterations.
The PSNRs for the left and right images are 9.0 dB, 9.4 dB, and  27.1 dB, respectively.

\section{Conclusion}
We proposed an iterative technique for generating amplitude CGHs and kinoforms using the random phase-free method with virtual convergence light.
The proposed method succeeded in mitigating ringing artifacts in the reconstructed images.

\section*{Acknowlegement}
This work is partially supported by JSPS KAKENHI Grant Numbers 25330125 and 25240015, and the Kayamori Foundation of Information Science Advancement and Yazaki Memorial Foundation for Science and Technology.

\bibliographystyle{model1a-num-names}
\bibliography{<your-bib-database>}

\end{document}